\documentclass[12pt,tightenlines,eqsecnum,floats,aps,amsmath,amssymb,nofootinbib,prd,showpacs]{revtex4}

\usepackage{setspace}
\usepackage{amsmath,amssymb,amsfonts,amsthm}
\usepackage{graphicx}
\usepackage{enumerate}
\usepackage{mathrsfs}

\begin{document}

\title{Schr\"{o}dinger quantization of linearly polarized Gowdy $\mathbb{S}^{1}\times\mathbb{S}^{2}$ and $\mathbb{S}^{3}$ models coupled to massless scalar fields}

\author{Daniel  \surname{G\'omez Vergel}}
\email[]{dgvergel@iem.cfmac.csic.es} \affiliation{Instituto de
Estructura de la Materia, CSIC, Serrano 121, 28006 Madrid, Spain}

\date{February 21, 2008}

\begin{abstract}
\indent In this paper we will construct the Schr\"{o}dinger
representation for the linearly polarized Gowdy
$\mathbb{S}^{1}\times\mathbb{S}^{2}$ and $\mathbb{S}^{3}$ models
coupled to massless scalar fields. Here the quantum states belong to
a $L^2$-space for a suitable quantum configuration space endowed
with a Gaussian measure, whose support is analyzed. This study
completes the quantization of these systems previously performed in
the Fock scheme, and provides a specially useful framework to
address physically relevant questions.
\end{abstract}


\maketitle

\section{Introduction}

\indent Gowdy models are $U(1)\times U(1)$ symmetry reductions with
many interesting applications in cosmology and quantum gravity,
since they provide inhomogeneous systems with local degrees of
freedom and invariance under a certain class of diffeomorphisms
\cite{Gowdy:1971jh}.

\indent The exact quantization of the linearly polarized Gowdy
$\mathbb{T}^3$ model in the vacuum has been profusely analyzed
\cite{Misner,Berger:1973,Mena:1997,Pierri:2000ri,
Corichi:2002vy,Torre:2002xt,Cortez,Corichi:2006xi,BarberoG.:2006zw,Corichi:2006zv,Mena:2007,
Torre:2007zj,Mena:2007schr}. Its gravitational local degrees of
freedom can be interpreted as those corresponding to a massless
scalar field in a fiducial background, so that the usual techniques
of QFT in curved spacetimes can be applied in order to construct the
quantum theory. The fact that the linear symplectic transformations
describing the classical time evolution cannot be unitarily
implemented in the physical Hilbert space when the system is written
in terms of its original variables was initially interpreted as a
serious obstacle for the feasibility of the model
\cite{Corichi:2002vy}. Nevertheless, it is possible to overcome this
problem by a suitable time-dependent redefinition of the field
\cite{Corichi:2006xi}. Furthermore, by demanding the unitarity of
the dynamics and the invariance under an extra $U(1)$ symmetry
generated by a residual global constraint, the existence of a unique
(up to unitary equivalence) Fock representation can be proved for
the system \cite{Corichi:2006zv,Mena:2007}.

\bigskip

\indent The existing literature has been recently extended to the
remaining topologies, $\mathbb{S}^{1}\times\mathbb{S}^{2}$ and
$\mathbb{S}^{3}$, allowing the coupling of gravity to massless
scalar fields (see \cite{BarberoG.:2007} for a rigorous classical
treatment of these models). Here, both gravitational and matter
local degrees of freedom can be encoded by massless scalar fields
evolving in the same fixed background metric. Therefore, they can be
treated in a unified way for the construction of the quantum theory.
A re-scaling of the fields similar to the one defined in the
three-torus case permits also a unitary implementation of the
dynamics \cite{BarberoG.:2007qt}. Concretely, this redefinition is
dictated by the conformal factor $\sin t$ that relates the Gowdy
metrics to the Einstein static (1+2)-universe. For these models, at
variance with the three-torus case, there is no extra constraint, so
that one obtains a family of (in general) unitarily nonequivalent
Fock representations, and in principle there is no symmetry argument
to select a preferred one. However, the uniqueness of the
representation can be recovered in these cases by imposing the
unitarity of the dynamics and the $SO(3)$ invariance of the Fock
construction\footnote{See also \cite{MenaSchr} for an independent
proof of this result. In this reference, some problems concerning
the completeness of the results given in \cite{BarberoG.:2007qt}
were pointed out. Nevertheless, they can be easily solved by
introducing some minor changes that will be taken into account in
the next section.} \cite{BarberoG.:2007qt}. Furthermore, it is
expected that a discussion similar to the one developed in
\cite{Mena:2007} for the vacuum Gowdy $\mathbb{T}^{3}$ model will
lead us to conclude that this redefinition of the fields is the only
reasonable one (up to multiplicative constants) providing unitary
dynamics under the condition of $SO(3)$ invariance.

\bigskip

\indent We will consider the Schr\"{o}dinger representation for the
linearly polarized Gowdy $\mathbb{S}^{1}\times\mathbb{S}^{2}$ and
$\mathbb{S}^{3}$ models coupled to massless scalar fields, where the
states act as functionals on the quantum configuration space
$\overline{\mathscr{C}}$ for a fixed time $t_0$. Here
$\overline{\mathscr{C}}$ is an appropriate distributional extension
of the classical configuration space $\mathscr{C}$, taken in these
cases to be the space of tempered distributions on the 2-sphere. The
Hilbert space then takes the form
$\mathscr{H}_{s}(t_0)=L^{2}(\overline{\mathscr{C}},\mathrm{d}\mu_{t_0})$.
The identification of the Gaussian nature of the measure
$\mu_{t_0}$, the nonstandard representation of the momentum
operator, and the relation between Schr\"{o}dinger and Fock
representations were exhaustively analyzed in \cite{CorichiSchr1} as
a natural extension to the functional description of the Fock
quantization of scalar fields in curved backgrounds \cite{Wald}. In
the QFT context, the Schr\"{o}dinger representation has been
historically pushed into the background in favor of the usual Fock
one because of the difficulty in using it to address sensible
questions regarding physical scattering processes. However, it is
certainly the most natural representation in the context of
canonical quantum gravity, in view of the splitting of spacetime
into spatial sections of constant time. Furthermore, as was pointed
out in \cite{Torre:2007zj} for the vacuum three-torus case, it
provides a better understanding of the properties of the quantized
field, since it is possible to determine the behavior of the typical
field confi\-gurations through the study of the measure support. The
Schr\"{o}dinger representation is also a privileged framework to
probe the existence and properties of semiclassical and squeezed
states for these systems. This paper represents then a necessary
first step to tackle this issue that will be considered elsewhere
after the rigorous analysis of this type of quantizations
\cite{Preparation}. Note that this is not a trivial question owing
to the nonautonomous nature of the Hamiltonian that governs the
reduced dynamics of the models.

\bigskip

In this paper, we will closely rely on the notation and results of
\cite{BarberoG.:2007,BarberoG.:2007qt}, where the reader can find
the classical and quantum formulations of the Gowdy
$\mathbb{S}^{1}\times\mathbb{S}^{2}$ and $\mathbb{S}^{3}$ models, as
well as on the recent works \cite{Torre:2007zj,Mena:2007schr},
devoted to the Schr\"{o}dinger representation for the vacuum Gowdy
$\mathbb{T}^{3}$ model both for the original and the redefined
scalar fields, respectively. In particular, the need for extending
the results found for the three-torus case to the remaining
topologies, and discussing the differences between them, was already
pointed out in \cite{Torre:2007zj}. In section \ref{FOCK} we will
summarize the features of the Fock construction for the Gowdy
$\mathbb{S}^{1}\times\mathbb{S}^{2}$ and $\mathbb{S}^{3}$ models
corresponding to the re-scaled fields for which the dynamics is
unitary, analyzing in subsection \ref{Self-adjointness} the
implementation of the Hamiltonian as a self-adjoint operator for
each value of the time parameter. We will also discuss here the
possibility of modifying the expression of the Hamiltonian at the
classical level in order to avoid some problems regarding the domain
of its quantum counterpart. In section \ref{SCHROD}, we will proceed
to define the Schr\"{o}dinger representation for these models in
such a way that the construction is unitarily equivalent to the Fock
one. In particular, we will probe the properties and support of the
measure $\mu_{t_0}$ in subsection \ref{METRIC}, as well as the
representation of the canonical commutation relations in subsection
\ref{CCR}. In section \ref{UNITARY}, we will check that, as a
consequence of the unitary implementation of the time evolution, the
representations corresponding to different values of the time
parameter are unitarily equivalent, and also that their associated
measures are mutually absolutely continuous. Finally, in section
\ref{COMMENTS}, we will make some comments and remarks on the
results of the paper, in particular concerning their similarity with
those found for the three-torus case.

\section{Fock representation}\label{FOCK}

\subsection{General framework}

\indent The dynamics of both gravitational and matter local degrees
of freedom\footnote{In what follows we will consider the use of the
redefined scalar field $\xi$ for which the dynamics can be unitarily
implemented \cite{BarberoG.:2007qt}. We will not study the global
modes present in these models \cite{BarberoG.:2007}. They can be
quantized in a straightforward way in terms of standard position and
momentum operators with dense domain in $L^{2}(\mathbb{R})$.} in the
linearly polarized Gowdy $\mathbb{S}^{1}\times\mathbb{S}^{2}$ and
$\mathbb{S}^{3}$ models can be described by the same non\-autonomous
Hamiltonian system $(\mathbf{P},\omega,H(t))$, whose features we
proceed to summarize. Let
$\gamma_{ab}=(\mathrm{d}\theta)_{a}(\mathrm{d}\theta)_{b}+\sin^{2}\theta(\mathrm{d}\sigma)_{a}(\mathrm{d}\sigma)_{b}\,$
be the round metric in the 2-sphere $\mathbb{S}^{2}$, with spherical
coordinates $(\theta,\sigma)\in(0,\pi)\times(0,2\pi)$. $\mathbf{P}$
is the space of smooth and symmetric Cauchy data $(Q,P)\in
C^{\infty}(\mathbb{S}^{2};\mathbb{R})\times
C^{\infty}(\mathbb{S}^{2};\mathbb{R})$, with
$\mathcal{L}_{\sigma}Q=0=\mathcal{L}_{\sigma}P$, where
$\mathcal{L}_{\sigma}$ denotes the Lie derivative with respect to
the vector field $\sigma^{a}=(\partial/\partial\sigma)^{a}$. The
standard (weakly) symplectic structure
$\omega:\mathbf{P}\times\mathbf{P}\rightarrow\mathbb{R}$ is given by
\begin{equation}\label{omega}
\omega((Q_1,P_1),(Q_2,P_2)):=\int_{\mathbb{S}^{2}}|\gamma|^{1/2}(Q_{2}P_{1}-Q_{1}P_{2})\,,\,\,\,(Q_1,P_1),(Q_2,P_2)\in\mathbf{P}\,.
\end{equation}
The symplectic space $(\mathbf{P},\omega)$ is then the
\emph{canonical phase space} of the system. Finally,
$H:(0,\pi)\times\mathbf{P}\rightarrow\mathbb{R}$ is the (indefinite)
nonautonomous Hamiltonian
\begin{equation}\label{ClassHamiltonian}
H(t;Q,P):=\frac{1}{2}\int_{\mathbb{S}^{2}}|\gamma|^{1/2}(P^2+\cot
t\,QP-Q\Delta_{\mathbb{S}^{2}}Q)\,,
\end{equation}
where $\Delta_{\mathbb{S}^{2}}$ denotes the Laplace-Beltrami
operator on the round 2-sphere. Consider now the space of smooth and
symmetric real solutions to the Euler-Lagrange equation derived from
the Hamilton equations\footnote{The dot denotes time derivative.}
\begin{equation}
\mathcal{S}:=\Big\{\xi\in
C^{\infty}((0,\pi)\times\mathbb{S}^{2};\mathbb{R})\,\big|\,-\ddot{\xi}+\Delta_{\mathbb{S}^{2}}\xi=\frac{1}{4}(1+\csc^{2}t)\xi\,,\,\,\,\mathcal{L}_{\sigma}\xi=0\Big\}\,.
\end{equation}
We define the \emph{covariant phase space} of the system as the pair
$(\mathcal{S},\Omega)$, where
$\Omega:\mathcal{S}\times\mathcal{S}\rightarrow\mathbb{R}$ is the
symplectic structure naturally induced by the $\omega$ given in
(\ref{omega}),
\begin{equation}\label{Omega}
\Omega(\xi_1,\xi_2):=\int_{\mathbb{S}^{2}}|\gamma|^{1/2}\iota_{t}^{*}(\xi_2\dot{\xi}_1-\xi_1\dot{\xi}_2)\,,\,\,\,\xi_1,\xi_2\in\mathcal{S}\,,
\end{equation}
with
$\iota_{t}:\mathbb{S}^{2}\rightarrow(0,\pi)\times\mathbb{S}^{2}$
being the embedding of the 2-sphere as a Cauchy surface of constant
time $t$.

\bigskip

\indent In order to obtain the quantum theory for these models, it
is necessary to construct the one-particle Hilbert space of the
system $\mathscr{H}_{\mathcal{P}}$. Consider the Lagrangian subspace
\begin{equation}\label{subspaceP}
\mathcal{P}:=\Big\{Z\in\mathcal{S}_{\mathbb{C}}\,\big|\,Z=\sum_{\ell=0}^{\infty}a_{\ell}z_{\ell}Y_{\ell
0}\,,\,\,a_{\ell}\in\mathbb{C}\Big\}
\end{equation}
of the complexification $\mathcal{S}_{\mathbb{C}}$ of the solution
space $\mathcal{S}$, where $(z_{\ell})_{\ell=0}^{\infty}$ is a
family of complex linearly independent solutions to the equation
\begin{equation}\label{equationz}
\ddot{z}_{\ell}+\left(\frac{1}{4}\big(1+\csc^{2}t\big)+\ell(\ell+1)\right)z_{\ell}=0
\end{equation}
satisfying the normalization condition\footnote{The bar denotes
complex conjugation.}
\begin{equation}\label{norm}
z_{\ell}\dot{\bar{z}}_{\ell}-\bar{z}_{\ell}\dot{z}_{\ell}=i\,,
\end{equation}
and $Y_{\ell 0}$ are the spherical harmonics verifying the
orthogonality conditions $\int_{\mathbb{S}^{2}}|\gamma|^{1/2}Y_{\ell
0}Y_{\ell^{\prime}0}\\=\delta(\ell,\ell^{\prime})$. The one-particle
Hilbert space $\mathscr{H}_{\mathcal{P}}$ is then the Cauchy
completion of the subspace $\mathcal{P}$ with respect to the inner
product
\begin{equation}
\langle
Z_1\,|\,Z_2\rangle_{\mathcal{P}}:=-i\Omega_{\mathbb{C}}(\bar{Z}_1,Z_2)=\sum_{\ell=0}^{\infty}\bar{a}^{(1)}_{\ell}a^{(2)}_{\ell}\,,\,\,\,Z_{1},Z_{2}\in\mathcal{P}\,,
\end{equation}
where $\Omega_{\mathbb{C}}$ is the extension of the symplectic
structure (\ref{Omega}) to $\mathcal{S}_{\mathbb{C}}$ by linearity.
Finally, the Hilbert space\footnote{The unnecessary distinction
between kinematical and physical Hilbert spaces in these models
follows from the nonexistence of extra constraints
\cite{BarberoG.:2007}.} of the models is given by the symmetric Fock
space defined on $\mathscr{H}_{\mathcal{P}}$,
\begin{equation}
\mathscr{F}_{\mathcal{P}}:=\bigoplus_{n=0}^{\infty}\mathscr{H}_{\mathcal{P}}^{\otimes_{s}n}\,,
\end{equation}
where $\mathscr{H}^{\otimes_{s}n}_{\mathcal{P}}$ denotes the
subspace of $\mathscr{H}_{\mathcal{P}}^{\otimes
n}=\otimes_{k=1}^{n}\mathscr{H}_{\mathcal{P}}$ spanned by symmetric
tensor pro\-ducts of $n$ vectors in $\mathscr{H}_{\mathcal{P}}$. The
possible choices of Lagrangian subspaces $\mathcal{P}$ are encoded
in the following two-parameter family of $z_{\ell}$ functions
satisfying (\ref{norm}):
\begin{equation}
z_{\ell}(t)=\sqrt{\frac{\sin
t}{2}}\left(\rho_{\ell}\mathscr{P}_{\ell}(\cos
t)+(\nu_{\ell}+i\rho_{\ell}^{-1})\mathscr{Q}_{\ell}(\cos t)\right),
\end{equation}
with $\rho_{\ell}>0$, $\nu_{\ell}\in\mathbb{R}$, modulo a
multiplicative phase that plays no role in the context of the study
of unitary implementation of dynamics. $\mathscr{P}_{\ell}$ and
$\mathscr{Q}_{\ell}$ denote the first and second class Legendre
functions, respectively.\\
\indent Every election of $\mathcal{P}$ is in one-to-one
correspondence with a $\Omega$-compatible $SO(3)$-invariant complex
structure on $\mathcal{S}$,
$J_{\mathcal{P}}:\mathcal{S}\rightarrow\mathcal{S}$,
$J_{\mathcal{P}}^{2}=-\mathrm{Id}_{\mathcal{S}}$ (see
\cite{BarberoG.:2007qt}). Indeed, any solution $\xi\in\mathcal{S}$
can be uniquely decomposed as $\xi=Z+\bar{Z}$, with
$Z\in\mathcal{P}$, in such a way that $J_\mathcal{P}$ is defined as
\begin{equation}\label{Jp}
J_{\mathcal{P}}\xi:=i(Z-\bar{Z})\,.
\end{equation}
As proved in \cite{BarberoG.:2007qt}, the linear symplectic
transformations that describe the time evolution can be unitarily
implemented in the Hilbert space $\mathscr{F}_{\mathcal{P}}$ for all
those $SO(3)$-invariant complex structures $J_{\mathcal{P}}$
characterized by pairs $(\rho_{\ell},\nu_{\ell})_{\ell=0}^{\infty}$
such that\footnote{The expression of the $\rho_{\ell}$ coefficients
appearing in \cite{BarberoG.:2007qt} is incomplete, and needs to be
corrected by taking into account the subdominant term that appears
in (\ref{rhonu}). With more generality, the expression of
$\nu_{\ell}$ given in \cite{BarberoG.:2007qt} must also be replaced
by the one of equation (\ref{rhonu}) in order to explicitly include
nonpolynomial decreasing behaviors. Taking these minor changes into
consideration, we completely characterize the biparametric family of
complex structures for which dynamics is unitary, and not only a
subfamily as in \cite{BarberoG.:2007qt}, solving the problems
pointed out at the end of reference \cite{MenaSchr}. We must remark,
in any case, that these corrections do not affect the main
conclusions achieved in \cite{BarberoG.:2007qt}, in particular,
concerning the proof of uniqueness of the Fock representation, whose
simplicity typifies the usefulness of the formalism developed in
\cite{BarberoG.:2007qt}.}
\begin{equation}\label{rhonu}
\rho_{\ell}=\sqrt{\frac{\pi}{2}}+x_{\ell}>0\,,\,\,\,(x_{\ell})_{\ell=0}^{\infty}\in\ell^{2}(\mathbb{R})\,,\quad\mathrm{and}\quad
(\nu_{\ell})_{\ell=0}^{\infty}\in\ell^{2}(\mathbb{R})\,.
\end{equation}
In addition, all the Fock representations obtained through
(\ref{rhonu}) are unitarily equivalent
\cite{BarberoG.:2007qt,MenaSchr}. In the following, we will
implicitly assume the use of a concrete complex structure
$J_{\mathcal{P}}$ of this type.

\subsection{Self-adjointness of the quantum
Hamiltonian}\label{Self-adjointness}

\indent Note that due to the nonautonomous nature of the classical
Hamiltonian (\ref{ClassHamiltonian}), the dynamics does not define a
one-parameter symplectic group on $(\mathbf{P},\omega)$, so we
cannot apply Stone's theorem to justify the self-adjointness of the
corresponding (one-parameter family of) operators in the quantum
theory. Nevertheless, it is possible to show that the quantum
Hamiltonian is self-adjoint for each value of the time parameter $t$
by analyzing the unitary implementability on
$\mathscr{F}_{\mathcal{P}}$ of the one-parameter symplectic group
generated by the \emph{autonomous} Hamiltonian $H(\tau)$, once a
value $t=\tau\in(0,\pi)$ has been fixed. Here, we will follow the
efficient procedure employed in \cite{Torre:2002xt} for the Gowdy
$\mathbb{T}^{3}$ model, subsequently genera\-lized in
\cite{BarberoG.:2006zw} to discuss the self-adjointness of general
quadratic operators in this context. We start by considering the
auxiliary system $(\mathbf{P},\omega,H(\tau))$, where the dynamics
is governed by the classical autonomous Hamiltonian
\begin{equation}
H(\tau)=\frac{1}{2}\sum_{\ell=0}^{\infty}\big(K_{\ell}(\tau)a_{\ell}^{2}+\bar{K}_{\ell}(\tau)\bar{a}_{\ell}^{2}+2G_{\ell}(\tau)\bar{a}_{\ell}a_{\ell}\big)\,,
\end{equation}
with
\begin{eqnarray}\label{K&G}
K_{\ell}(\tau)&:=&\Big(\dot{z}_{\ell}(\tau)-\frac{1}{2}\cot
\tau\,z_{\ell}(\tau)\Big)^{2}+\ell(\ell+1)z_{\ell}^{2}(\tau)+\cot
\tau\,\Big(\dot{z}_{\ell}(\tau)-\frac{1}{2}\cot
\tau\,z_{\ell}(\tau)\Big)z_{\ell}(\tau)\,,\nonumber\\
G_{\ell}(\tau)&:=&\Big|\dot{z}_{\ell}(\tau)-\frac{1}{2}\cot
\tau\,z_{\ell}(\tau)\Big|^{2}+\ell(\ell+1)|z_{\ell}(\tau)|^{2}\nonumber\\
&&+\frac{1}{2}\cot\tau\,\left(\Big(\dot{z}_{\ell}(\tau)-\frac{1}{2}\cot
\tau\,z_{\ell}(\tau)\Big)\bar{z}_{\ell}(\tau)+\Big(\dot{\bar{z}}_{\ell}(\tau)-\frac{1}{2}\cot
\tau\,\bar{z}_{\ell}(\tau)\Big)z_{\ell}(\tau)\right).
\end{eqnarray}
The modes $a_{\ell},\bar{a}_{\ell}$ are defined through the
relations
$Q_{\ell}:=\int_{\mathbb{S}^{2}}|\gamma|^{1/2}QY_{\ell0}=z_{\ell}(\tau)a_{\ell}+\bar{z}_{\ell}(\tau)\bar{a}_{\ell}$,
$P_{\ell}:=\int_{\mathbb{S}^{2}}|\gamma|^{1/2}PY_{\ell0}=\big(\dot{z}_{\ell}(\tau)-(1/2)\cot\tau\,z_{\ell}(\tau)\big)a_{\ell}+\big(\dot{\bar{z}}_{\ell}(\tau)-(1/2)\cot\tau\,\bar{z}_{\ell}(\tau)\big)\bar{a}_{\ell}$.
Their evolution in a fictitious time parameter $s\in\mathbb{R}$ is
given by the linear equations\footnote{Here $\{\cdot,\cdot\}$
denotes the Poisson bracket defined from (\ref{omega}), with
$\{a_{\ell},\bar{a}_{\ell^{\prime}}\}=-i\delta(\ell,\ell^{\prime})\mathbb{I}$.}
\begin{eqnarray}
\frac{\mathrm{d}a_{\ell}}{\mathrm{d}s}&=&\{a_{\ell},H(\tau)\}=-i\big(G_{\ell}(\tau)a_{\ell}+\bar{K}_{\ell}(\tau)\bar{a}_{\ell}\big)\,,\\
\frac{\mathrm{d}\bar{a}_{\ell}}{\mathrm{d}s}&=&\{\bar{a}_{\ell},H(\tau)\}=i\big(K_{\ell}(\tau)a_{\ell}+G_{\ell}(\tau)\bar{a}_{\ell}\big)\,.\nonumber
\end{eqnarray}
Using the normalization condition (\ref{norm}), we easily obtain the
second-order differential equation
\begin{equation}\label{dderiva}
\frac{\mathrm{d}^{2}a_{\ell}}{\mathrm{d}s^{2}}=-\Big(\ell(\ell+1)-\frac{1}{4}\cot^{2}\tau\Big)a_{\ell}\,,
\end{equation}
whose solutions have a linear dependence on the initial conditions
$a_{\ell}(s_0)$ and $\bar{a}_{\ell}(s_0)$,
\begin{eqnarray}\label{transforma}
a_{\ell}(s)=\alpha_{\ell}(s,s_0)a_{\ell}(s_0)+\beta_{\ell}(s,s_0)\bar{a}_{\ell}(s_0)\,,\,\,\,\bar{a}_{\ell}(s)=\overline{a_{\ell}(s)}\,.
\end{eqnarray}
This symplectic transformation is unitarily implementable on
$\mathscr{F}_{\mathcal{P}}$ for each $s\in\mathbb{R}$, i.e., there
exists a unitary operator
$\hat{u}(s,s_0):\mathscr{F}_{\mathcal{P}}\rightarrow\mathscr{F}_{\mathcal{P}}$
such that
$\hat{u}(s,s_0)\hat{a}_{\ell}\hat{u}^{-1}(s,s_0)=\alpha_{\ell}(s,s_0)\hat{a}_{\ell}+\beta_{\ell}(s,s_0)\hat{a}_{\ell}^{\dag}$,
$\hat{u}(s,s_0)\hat{a}_{\ell}^{\dag}\hat{u}^{-1}(s,s_0)=\bar{\beta}_{\ell}(s,s_0)\hat{a}_{\ell}+\bar{\alpha}_{\ell}(s,s_0)\hat{a}_{\ell}^{\dag}$,
if and only if the Bogoliubov coefficients $\beta_{\ell}$ are square
summable \cite{Shale},
\begin{equation}\label{betacoeff}
\sum_{\ell=0}^{\infty}|\beta_{\ell}(s,s_0)|^{2}<+\infty\,.
\end{equation}
Note that, for each value of $\tau\in(0,\pi)$, there exists
$\ell_{0}\in\mathbb{N}\cup\{0\}$ such that
\begin{equation*}
\lambda_{\ell}^{2}:=\ell(\ell+1)-\frac{1}{4}\cot^{2}\tau>0\,,\,\,\,\forall\,\ell>\ell_{0}\,.
\end{equation*}
In this situation,
\begin{eqnarray*}
\alpha_{\ell}(s,s_0)&=&\cos\big(\lambda_{\ell}(s-s_0)\big)-i\lambda_{\ell}^{-1}G_{\ell}(\tau)\sin\big(\lambda_{\ell}(s-s_0)\big)\,,\\
\beta_{\ell}(s,s_0)&=&-i\lambda_{\ell}^{-1}\bar{K}_{\ell}(\tau)\sin\big(\lambda_{\ell}(s-s_0)\big)\,.
\end{eqnarray*}
It suffices to consider the modes corresponding to $\ell>\ell_{0}$,
since the convergence of the series (\ref{betacoeff}) depends, in
practice, only on the high-frequency behavior of the $\beta_{\ell}$
coefficients. Taking into account the asymptotic expansions in
$\ell$
\begin{eqnarray}\label{asymp}
&&z_{\ell}(t)=\frac{1}{\sqrt{2\ell}}\exp\left(-i[(\ell+1/2)t-\pi/4]\right)+O(\ell^{-3/2})\,,\\
&&\dot{z}_{\ell}(t)-\frac{1}{2}\cot t\,z_{\ell}(t)=
-i\sqrt{\frac{\ell}{2}}\exp\left(-i[(\ell+1/2)t-\pi/4]\right)+O(\ell^{-1/2})\,,\nonumber
\end{eqnarray}
we have $K_{\ell}(\tau)=O(1)$, so that
$\sum_{\ell>\ell_0}\lambda_{\ell}^{-2}|K_{\ell}(\tau)|^{2}\sin^{2}\big(\lambda_{\ell}(s-s_0)\big)<+\infty$,
$\forall\,s\in\mathbb{R}$, and hence (\ref{betacoeff}) is verified.
Finally, the transformation (\ref{transforma}) is implementable as a
continuous, unitary, one-parameter group if it verifies the strong
continuity condition in the auxiliary parameter $s$
\begin{equation}
\lim_{s\rightarrow
s_{0}}\sum_{\ell=0}^{\infty}|a_{\ell}(s)-a_{\ell}(s_0)|^{2}=0\,,\,\,\,s_{0}\in\mathbb{R}\,.
\end{equation}
Again, we can restrict ourselves to the modes $\ell>\ell_0$. It is
straightforward to check that this condition holds for the solution
(\ref{transforma}) with square summable initial data $a_{\ell}$ and
$\bar{a}_{\ell}$. Therefore, we have obtained a strongly continuous
and unitary one-parameter group whose generator is self-adjoint
according to Stone's theorem.

\bigskip

The quantum Hamiltonian of the models under consideration can be
explicitly calculated as the strong limit
\begin{equation*}
\mathrm{s-}\!\!\lim_{t^{\prime}\rightarrow
t}\frac{\hat{U}(t,t^{\prime})-\hat{\mathbb{I}}}{t-t^{\prime}}f=-i\hat{H}(t)f\,,\,\,\,
f\in\mathscr{D}_{\hat{H}(t)}\,,
\end{equation*}
where $\hat{U}(t,t^{\prime})$ denotes the quantum evolution operator
on $\mathscr{F}_{\mathcal{P}}$. The previous result ensures the
self-adjointness of the quantum Hamiltonian $\hat{H}(t)$, and hence
the existence of a dense domain
$\mathscr{D}_{\hat{H}(t)}\subset\mathscr{F}_{\mathcal{P}}$, for each
value of the time parameter $t\in(0,\pi)$. Unfortunately, the method
employed does not provide us with a characterization of such
domains, or the concrete expression of the quantum Hamiltonian.
Nevertheless, given the quadratic nature of the classical
Hamiltonian (\ref{ClassHamiltonian}), it is expected that this limit
coincides with the operator directly promoted from the classical
function up to normal ordering. As proved in
\cite{BarberoG.:2007qt}, this operator does not have the Fock vacuum
state $|0\rangle_{\mathcal{P}}:=1\oplus
0\oplus0\oplus\cdots\in\mathscr{F}_{\mathcal{P}}$ in its domain
because of the fact that the $K_{\ell}$ sequence defined in
(\ref{K&G}) is not square summable. As a consequence, the action of
the operator is not defined either on the dense subspace of states
with a finite number of particles. This difficulty can be overcome
right from the start by describing the classical dynamics through
the (positive definite) Hamiltonian \cite{BarberoG.:2007qt}
\begin{equation}\label{NewClassicHamilt}
H_{0}(Q,P;t):=\frac{1}{2}\int_{\mathbb{S}^{2}}|\gamma|^{1/2}\left(P^{2}+Q\Big[\frac{1}{4}(1+\csc^{2}t)-\Delta_{\mathbb{S}^{2}}\Big]Q\right).
\end{equation}
The Hamiltonians (\ref{ClassHamiltonian}) and
(\ref{NewClassicHamilt}) obviously govern the same classical
evolution, but they are connected by a time-dependent symplectic
transformation that in principle is not unitarily implementable, so
one possibly obtains nonequivalent quantum theories from them. The
corresponding quantum Hamiltonian is given, after normal ordering,
by
\begin{equation}\label{QuantHamiltonian}
\hat{H}_{0}(t)=\frac{1}{2}\sum_{\ell=0}^{\infty}\left(K_{0_{\ell}}(t)\hat{a}_{\ell}^{2}+\bar{K}_{0_\ell}(t){\hat{a}_{\ell}}^{\dag
2}+2G_{0_\ell}(t)\hat{a}_{\ell}^{\dag}\hat{a}_{\ell}\right),
\end{equation}
where
\begin{eqnarray}\label{K0&G0}
K_{0_\ell}(t)&:=&\dot{z}_{\ell}^{2}(t)+\left(\frac{1}{4}\big(1+\csc^{2}t\big)+\ell(\ell+1)\right)z_{\ell}^{2}(t)\,,\\
G_{0_\ell}(t)&:=&\big|\dot{z}_{\ell}(t)\big|^{2}+\left(\frac{1}{4}\big(1+\csc^{2}t\big)+\ell(\ell+1)\right)|z_{\ell}(t)|^{2}\,.\nonumber
\end{eqnarray}
Here, $\hat{a}^{\dag}_{\ell}$ and $\hat{a}_{\ell}$ are the creation
and annihilation operators associated with the modes
$z_{\ell}Y_{\ell0}$, respectively. This new self-adjoint Hamiltonian
has the advantage of including the vacuum state in its domain --in
this case $K_{0_\ell}(t)$ defines a square summable sequence for
each value of $t$--, as well as the fact that the results about the
unitary implementation of the time evolution and the uniqueness of
the Fock representation are also valid in this case. Concretely, the
biparametric family of complex structures for which the dynamics is
unitary is characterized again by the pairs (\ref{rhonu}). In what
follows, we will consider the dynamics of the system to be described
by (\ref{NewClassicHamilt}).

\section{Schr\"{o}dinger representation}\label{SCHROD}

\subsection{Constructing the $L^2$ space}

\indent Let us denote by $\mathscr{S}$ the Schwartz space of smooth
and symmetric test functions on the 2-sphere,
\begin{equation}
\mathscr{S}:=\{f\in
C^{\infty}(\mathbb{S}^{2};\mathbb{R})\,\,|\,\,\mathcal{L}_{\sigma}f=0\}\,,
\end{equation}
endowed with the standard nuclear
topology\footnote{\label{footnote}Every element $f\in\mathscr{S}$
can be expanded as $f(s)=\sum_{\ell=0}^{\infty}f_{\ell}Y_{\ell
0}(s)$, $s\in\mathbb{S}^{2}$, with $(f_\ell)_{\ell=0}^{\infty}$ a
sequence of rapidly decreasing real coefficients, such that
$\lim_{\ell\rightarrow\infty}\ell^{n}f_{\ell}=
0,\,\forall\,n\in\mathbb{N}\cup\{0\}$. We will revise the equivalent
description of the topological structure of $\mathscr{S}$ in terms
of the locally convex space of rapidly decreasing sequences in
section \ref{METRIC}. For more details, the reader can consult
\cite{Dubin&Hennings}.}. The quantum configuration space used to
define the Schr\"{o}dinger representation is then the topological
dual $\mathscr{S}^{\prime}$, consisting of conti\-nuous linear
functionals on $\mathscr{S}$. Note that this space includes the
delta functions and their derivatives. Given a time of embedding
$t_0$, the Schr\"{o}dinger representation is introduced by defining
a suitable Hilbert space\footnote{Here, the measure $\mu_{t_0}$ is
implicitly assumed to be defined on the sigma algebra
$\sigma(\mathrm{Cyl}(\mathscr{S}^{\prime}))$ generated by the
cylinder sets.} $L^{2}(\mathscr{S}^{\prime},\mathrm{d}\mu_{t_0})$,
for a certain measure $\mu_{t_0}$, in which the configuration
observables act as \emph{multiplication} operators. As we will see
later, given the Gaussian nature of the measure $\mu_{t_0}$, the
momentum operators will differ from the usual ones in terms of
derivatives by a multiplicative term depending on the configuration
variables.

\bigskip

\indent As a consequence of the linearity of
$\mathbf{P}=\mathscr{S}\times\mathscr{S}$, the set of elementary
classical observables $\mathcal{O}$ can be identified with the
$\mathbb{R}$-vector space generated by linear functionals on
$\mathbf{P}$. Every pair $\lambda:=(-g,f)\in\mathbf{P}$,
$f,g\in\mathscr{S}$, has an associated functional
$F_{\lambda}:\mathbf{P}\rightarrow\mathbb{R}$ such that for all
$X=(Q,P)\in\mathbf{P}$,
\begin{equation}
F_{\lambda}(X):=\omega(\lambda,X)=\int_{\mathbb{S}^{2}}|\gamma|^{1/2}(fQ+gP)\,.
\end{equation}
Therefore,
$\mathcal{O}=\mathrm{Span}\{\mathbb{I},F_{\lambda}\}_{\lambda\in\mathbf{P}}$.
As expected \cite{Ashtekar1}, this set satisfies the condition that
any regular function on $\mathbf{P}$ can be obtained as a (suitable
limit of) sum of products of elements in $\mathcal{O}$, and also
that it is closed under Poisson brackets,
$\{F_{\lambda}(\cdot),F_{\nu}(\cdot)\}=F_{\nu}(\lambda)\mathbb{I}$.
The configuration and momentum observables are objects of this type
defined by the pairs $\lambda=(0,f)$ and $\lambda=(-g,0)$,
respectively
\begin{eqnarray}
Q(f)&:=&F_{(0,f)}(Q,P)=\int_{\mathbb{S}^{2}}|\gamma|^{1/2}fQ=\sum_{\ell=0}^{\infty}f_{\ell}Q_{\ell}\,,\\
P(g)&:=&F_{(-g,0)}(Q,P)=\int_{\mathbb{S}^{2}}|\gamma|^{1/2}gP=\sum_{\ell=0}^{\infty}g_{\ell}P_{\ell}\,,
\end{eqnarray}
where the symmetric test functions have been expanded as explained
in footnote \ref{footnote}. Here, with the aim of simplifying the
notation, we have used the same symbol to denote the canonical
inclusion $\mathscr{S}\hookrightarrow\mathscr{S}^{\prime}$ of
$\mathscr{S}$ into $\mathscr{S}^{\prime}$. In this way,
$F_{(-g,f)}(Q,P)=Q(f)+P(g)$. The abstract quantum algebra of
observables $\mathcal{A}$ is then given by the usual Weyl
$C^{*}$-algebra generated by the elements
$W(\lambda)=\exp(iF_{\lambda})$, $\lambda\in\mathbf{P}$, satisfying
the conditions
\begin{equation}
\displaystyle W(\lambda)^{*}=W(-\lambda)\,,\quad
W(\lambda_1)W(\lambda_2)=\mathrm{e}^{\frac{i}{2}\omega(\lambda_1,\lambda_2)}W(\lambda_1+\lambda_2)\,,
\end{equation}
containing the information about the canonical commutation
relations.

\bigskip

\indent Let $\mathfrak{I}_{t_0}:\mathbf{P}\rightarrow\mathcal{S}$,
$t_0\in(0,\pi)$, be the symplectomorphism that defines for each pair
of Cauchy data $(Q,P)\in\mathbf{P}$ the unique solution
$\xi\in\mathcal{S}$ such that, under the evolution given by the
Hamiltonian (\ref{NewClassicHamilt}), it satisfies
$\xi(t_0,s)=Q(s)$, $\dot{\xi}(t_0,s)=P(s)$. That is
\begin{equation}\label{I0QP}
\xi(t,s)=(\mathfrak{I}_{t_0}(Q,P))(t,s)=\sum_{\ell=0}^{\infty}\left(a_{\ell}(t_0)z_{\ell}(t)+\overline{a_{\ell}(t_0)z_{\ell}(t)}\right)Y_{\ell
0}(s)\in\mathcal{S}\,,
\end{equation}
with
\begin{equation}\label{alt0}
a_{\ell}(t_0):=i\bar{z}_{\ell}(t_0)P_{\ell}-i\dot{\bar{z}}_{\ell}(t_0)Q_{\ell}\,.
\end{equation}
This map gives rise to a natural $\omega$-compatible complex
structure on the canonical phase space given by
\begin{equation}
J_{t_0}:=\mathfrak{I}_{t_0}^{-1}\circ
J_{\mathcal{P}}\circ\mathfrak{I}_{t_0}:\mathbf{P}\rightarrow\mathbf{P}\,,
\end{equation}
such that
\begin{equation*}
(Q,P)\in\mathbf{P}\mapsto
J_{t_0}(Q,P)=(A(t_0)Q+B(t_0)P,D(t_0)Q+C(t_0)P)\in\mathbf{P}\,,
\end{equation*}
where
$A(t_0),B(t_0),C(t_0),D(t_0):\mathscr{S}\rightarrow\mathscr{S}$ are
linear operators satisfying, in virtue of the $\omega$-compatibility
\cite{Ashtekar2}, the relations
\begin{eqnarray*}
\langle f,B(t_0)f^{\prime}\rangle=\langle
B(t_0)f,f^{\prime}\rangle\,,\,\,\,\,\langle
g,D(t_0)g^{\prime}\rangle=\langle
D(t_0)g,g^{\prime}\rangle\,,\,\,\,\,\langle
f,A(t_0)g\rangle=-\langle C(t_0)f,g\rangle\,,
\end{eqnarray*}
for all $f,g,f^{\prime},g^{\prime}\in\mathscr{S}$. Here, we have
denoted $\langle f,g\rangle:=\int_{\mathbb{S}^{2}}|\gamma|^{1/2}fg$.
Also, given the condition $J_{t_0}^{2}=-\mathrm{Id}_{\mathbf{P}}$,
and assuming $B(t_0)$ invertible, the $C(t_0)$ and $D(t_0)$
operators can be expressed in terms of the $A(t_0)$ and $B(t_0)$
operators through the relations $C(t_0)=-B^{-1}(t_0)A(t_0)B(t_0)$
and $D(t_0)=-B^{-1}(t_0)(\mathbf{1}+A^{2}(t_0))$, respectively, in
such a way that the complex structure $J_{t_0}$ is completely
characterized by $A(t_0)$ and $B(t_0)$. Using equations (\ref{Jp})
and (\ref{I0QP}), it is straightforward to obtain\footnote{Note that
the zero mode $\ell=0$ has been included into the spherical harmonic
expansion of the test functions. The $B(t_0)$ operator is well
defined even for this mode, ultimately as a consequence of equation
(\ref{equationz}) verified by the $z_{\ell}$ functions, where the
squared frequency is positive definite $\forall\,t\in(0,\pi)$ when
$\ell=0$.}
\begin{eqnarray}\label{A&B}
\big(A(t_0)Q\big)(s)&=&\sum_{\ell=0}^{\infty}\big(\dot{\bar{z}}_{\ell}(t_0)z_{\ell}(t_0)+\dot{z}_{\ell}(t_0)\bar{z}_{\ell}(t_0)\big)Q_{\ell}Y_{\ell 0}(s)\,,\\
\big(B(t_0)P\big)(s)&=&-2\sum_{\ell=0}^{\infty}|z_{\ell}(t_0)|^{2}P_{\ell}Y_{\ell
0}(s)\,.\nonumber
\end{eqnarray}
It is worth noting that, given the rapidly decreasing nature of the
sequences $(Q_{\ell})_{\ell=0}^{\infty}$ and
$(P_{\ell})_{\ell=0}^{\infty}$, as well as the asymptotic behavior
of the $z_{\ell}$ functions decaying like (\ref{asymp}), the
$A(t_0)$ and $B(t_0)$ operators are well defined on $\mathscr{S}$.
In addition, $B(t_0)$ has an inverse operator
$B^{-1}(t_0):\mathscr{S}\rightarrow\mathscr{S}$ given by
\begin{equation}\label{invB}
\big(B^{-1}(t_0)P\big)(s)=-\frac{1}{2}\sum_{\ell=0}^{\infty}|z_{\ell}(t_0)|^{-2}P_{\ell}Y_{\ell
0}(s)\,.
\end{equation}
\indent By definition, once a time of embedding $t_0$ is fixed, the
states in the Schr\"{o}dinger representation are characterized as
functionals $\Psi:\mathscr{S}^{\prime}\rightarrow\mathbb{C}$
belonging to a certain Hilbert space
$\mathscr{H}_{s}(t_0)=L^{2}(\mathscr{S}^{\prime},\mathrm{d}\mu_{t_0})$.
Due to the infinite dimensionality of the quantum configuration
space, it is not possible to define a Lebesgue-type measure
$\mu_{t_0}$, but rather a probability one\footnote{This is, a
measure satisfying
$\int_{\mathscr{S}^{\prime}}\mathrm{d}\mu_{t_0}=1$.}. This
representation is constructed in such a way that it is associated
with the state $\varpi_{t_0}:\mathcal{A}\rightarrow\mathbb{C}$ on
the Weyl algebra $\mathcal{A}$ whose action on the elementary
observables is given by \cite{CorichiSchr1,Wald}
\begin{equation}\label{statesigma}
\varpi_{t_0}(W(\lambda))=\exp\left(-\frac{1}{4}\omega(J_{t_0}(\lambda),\lambda)\right),\,\,\,\lambda\in\mathbf{P}\,.
\end{equation}
We will check in section \ref{UNITARY} that the Schr\"{o}dinger
representations corresponding to different values of the time
parameter are unitarily equivalent due to the unitary
implementability of the dynamics. We require that the configuration
observables are represented as \emph{multiplication} operators, so
that for $\lambda=(0,f)\in\mathbf{P}$,
\begin{equation}\label{Qoperator}
\displaystyle \pi_{s}(t_0)\cdot
W(\lambda)|_{\lambda=(0,f)}=\exp(i\hat{Q}_{t_0}[f])\,,\quad
\left(\hat{Q}_{t_0}[f]\Psi\right)[\tilde{Q}]=\tilde{Q}(f)\Psi[\tilde{Q}]\,,
\end{equation}
where $\tilde{Q}\in\mathscr{S}^{\prime}$ denotes a generic
distribution of $\mathscr{S}^{\prime}$ and $\tilde{Q}(f)$ gives the
usual pairing between $\mathscr{S}$ and $\mathscr{S}^{\prime}$,
$\Psi\in\mathscr{D}_{\hat{Q}_{t_0}[f]}\subset\mathscr{H}_{s}(t_{0})$
(the self-adjointness of the configuration and momentum operators
will be discussed in subsection \ref{CCR}), and
$\pi_{s}(t_0):\mathcal{A}\rightarrow
\mathscr{L}(\mathscr{H}_{s}(t_0))$ is the map from the Weyl algebra
$\mathcal{A}$ to the collection of bounded linear operators on
$\mathscr{H}_{s}(t_0)$. In this way, the measure $\mu_{t_0}$ is
Gaussian with covariance $\mathcal{C}(t_0):=-B(t_0)/2$, and thus its
Fourier transform is given by\footnote{This equation corresponds to
the expectation value (\ref{statesigma}) evaluated for
$\lambda=(0,f)$ that must coincide with the integral
$\int_{\mathscr{S}^{\prime}}\bar{\Psi}^{(t_0)}_{0}(\exp(i\hat{Q}_{t_0}[f])\Psi^{(t_0)}_{0})\,\mathrm{d}\mu_{t_0}$,
where $\Psi^{(t_0)}_{0}\in\mathscr{H}_{s}(t_0)$ is the normalized
vacuum state.}
\begin{equation}
\int_{\mathscr{S}^{\prime}}\mathrm{e}^{i\tilde{Q}(f)}\,\mathrm{d}\mu_{t_0}[\tilde{Q}]=\exp\left(\frac{1}{4}\langle
f,B(t_0)f\rangle\right).
\end{equation}
The covariance operator
$\check{\mathcal{C}}_{t_0}:\mathbf{P}\rightarrow\mathbb{R}$ is
defined as $\check{\mathcal{C}}_{t_0}(f,g):=\langle
f,\mathcal{C}(t_0)g\rangle$, $f,g\in\mathscr{S}$. Since
$|z_{\ell}(t_0)|^{2}$ is bounded and positive definite
$\forall\,t\in(0,\pi)$ and $\forall\,\ell\in\mathbb{N}\cup\{0\}$, it
follows that, as expected, $\check{\mathcal{C}}_{t_0}$ is a
nondegenerate positive definite and continuous bilinear form on the
topological vector space $\mathscr{S}$.

\subsection{Properties of the measure}\label{METRIC}

\indent In order to easily visualize the nature of the measure
$\mu_{t_0}$, note that upon restriction on any number of coordinate
directions in $\mathscr{S}^{\prime}$, say
$\tilde{Q}_{\ell}=\tilde{Q}(Y_{\ell 0})$, $\ell=0,1,\ldots,n$, we
obtain
\begin{equation}\label{restrict}
\displaystyle
\mathrm{d}\mu_{t_0}|_{(\tilde{Q}_{\ell})_{\ell=0}^{n}}=\prod_{\ell=0}^{n}\,\frac{1}{\sqrt{2\pi}}\,|z_{\ell}(t_0)|^{-1}\exp\left(-\frac{1}{2}|z_{\ell}(t_0)|^{-2}\tilde{Q}_{\ell}^{2}\right)\mathrm{d}\tilde{Q}_{\ell}\,.
\end{equation}
in terms of the Lebesgue measures $\mathrm{d}\tilde{Q}_{\ell}$
\cite{Glimm&Jaffe}.

\bigskip

\indent Now, we will prove that the support of the measure is
smaller than $\mathscr{S}^{\prime}$. Concretely, it is given by the
topological dual of the subspace of symmetric functions in the
Sobolev space $H^{\epsilon}(\mathbb{S}^{2})$ on the 2-sphere, for
any $\epsilon>0$. With this aim, we will use the Bochner-Minlos
theorem that plays a key role in the characterization of measures on
functional spaces, closely relying on the analysis developed in
\cite{Simon1}. We first point out that the space of test functions
$\mathscr{S}$ is topologically isomorphic to
$\varsigma=\bigcap_{r\in\mathbb{Q}}\varsigma_{r}$, where
\begin{equation}
\varsigma_{r}:=\Big\{f=(f_{\ell})_{\ell=0}^{\infty}\,\,\big|\,\,\|f\|_{r}^{2}:=\sum_{\ell=0}^{\infty}(\ell+1/2)^{2r}f_{\ell}^{2}<+\infty\Big\}\,,
\end{equation}
endowed with the Fr\'{e}chet topology induced by the norms
$(\|\cdot\|_{r})_{r\in\mathbb{Q}}$. As a consequence of the
Bochner-Minlos theorem (see the theorem 2.3 of \cite{Simon1}), if
the covariance $\check{\mathcal{C}}_{t_0}$ is continuous in the norm
associated with some $\varsigma_{r}$, then the Gaussian measure
$\mu_{t_0}$ has support on any set of the form
\begin{equation}\label{setr}
\Big\{f\,\,\big|\,\,\sum_{\ell=0}^{\infty}(\ell+1/2)^{-2r-1-2\epsilon}f_{\ell}^{2}<+\infty\,,\,\,\epsilon>0\Big\}\subset\displaystyle\bigcup_{r\in\mathbb{Q}}\varsigma_{r}=\varsigma^{\prime}\,,
\end{equation}
where $\varsigma^{\prime}$ is the topological dual\footnote{Here,
$g\in\varsigma^{\prime}$ is associated with the linear functional
$L_{g}(f):=\sum_{\ell=0}^{\infty}f_{\ell}g_{\ell}$,
$f\in\varsigma$.} of $\varsigma$. In particular, given the
asymptotic behavior of the $z_{\ell}$ functions, it is
straightforward to check the continuity in the norm corresponding to
$r=-1/2$, i.e.,
\begin{equation}
\langle f,\mathcal{C}(t_0)f\rangle\le
N(t_0)\sum_{\ell=0}^{\infty}(\ell+1/2)^{-1}f_{\ell}^{2}
\end{equation}
for certain constant $N(t_0)\in\mathbb{R}^{+}$. According to this
result, the measure $\mu_{t_0}$ is concentrated on the set
(\ref{setr}) for $r=-1/2$, which can be identified with the
topological dual $\mathfrak{h}_{\epsilon}^{\prime}$ of the subspace
of symmetric functions in the Sobolev space
$H^{\epsilon}(\mathbb{S}^{2})$, for any $\epsilon>0$,
\begin{equation}
\mathfrak{h}_{\epsilon}:=\Big\{f\in
H^{\epsilon}(\mathbb{S}^{2})\,\,\big|\,\,\mathcal{L}_{\sigma}f=0\,,\,\,\|f\|_{\epsilon}^{2}:=\sum_{\ell=0}^{\infty}(\ell+1/2)^{2\epsilon}f_{\ell}^{2}<+\infty\Big\}\,,\,\,\,\epsilon>0\,,
\end{equation}
where $f_\ell$ are the Fourier coefficients of the function $f$.
Therefore, the typical field configurations are not as singular as
the delta functions or their derivatives. However, the subset
$\mathfrak{b}\subset\mathfrak{h}_{\epsilon}^{\prime}$ of symmetric
$L^{2}(\mathbb{S}^{2})$ functions has also measure zero. Indeed,
consider the characteristic function $\chi_{\mathfrak{b}}$ of the
measurable set $\mathfrak{b}$, defined by
\begin{equation}
\chi_{\mathfrak{b}}[\tilde{Q}]:=\lim_{\alpha\rightarrow
+0}\,\exp\left(-\alpha\sum_{\ell=0}^{\infty}\tilde{Q}_{\ell}^{2}\right),
\end{equation}
so that $\chi_{\mathfrak{b}}[\tilde{Q}]=1$, for $\tilde{Q}\in
\mathfrak{b}$, and vanishes anywhere else. Making use of the
restriction (\ref{restrict}), and applying the Lebesgue monotone
convergence theorem, it is straightforward to obtain
\begin{equation}
\mu_{t_0}(\mathfrak{b})=\int_{\mathscr{S}^{\prime}}\chi_{\mathfrak{b}}[\tilde{Q}]\,\mathrm{d}\mu_{t_0}[\tilde{Q}]=\lim_{\alpha\rightarrow
+0}\,\lim_{n\rightarrow\infty}\,\prod_{\ell=0}^{n}\frac{1}{\sqrt{1+2\alpha|z_{\ell}(t_0)|^{2}}}\,.
\end{equation}
The limit of the product vanishes as $n\rightarrow\infty$ because of
the nonconvergence of the series
$\sum_{\ell=0}^{\infty}\log(1+2\alpha|z_{\ell}(t_0)|^{2})$, and
hence\footnote{Since $\mathscr{S}\hookrightarrow\mathfrak{b}$, we
have that, as usual for a field theory, the measure $\mu_{t_0}$ is
not supported on the classical configuration space $\mathscr{S}$.
This is precisely the reason why a suitable distributional extension
of $\mathscr{S}$ must be chosen as measure space in order to
construct the $L^{2}$ space for the Schr\"{o}dinger representation.}
$\mu_{t_0}(\mathfrak{b})=0$.

\subsection{Canonical commutation relations}\label{CCR}

\indent By virtue of the interrelation between operator
representation and measures, the representation of the basic
momentum observables is \cite{CorichiSchr1}
\begin{eqnarray}
&&\pi_{s}(t_0)\cdot
W(\lambda)|_{\lambda=(-g,0)}=\exp(i\hat{P}_{t_0}[g])\,,\nonumber\\
&&\left(\hat{P}_{t_0}[g]\Psi\right)[\tilde{Q}]=-i(D_{\tilde{Q}}\Psi)[g]-i\tilde{Q}\left(B^{-1}(t_0)(\mathbf{1}-i
A(t_0))g\right)\Psi[\tilde{Q}]\,,
\end{eqnarray}
where $\tilde{Q}\in\mathscr{S}^{\prime}$,
$\Psi\in\mathscr{D}_{\hat{P}_{t_0}[g]}\subset\mathscr{H}_{s}(t_{0})$,
and $(D_{\tilde{Q}}\Psi)$ denotes the directional derivative of the
functional $\Psi$ in the direction defined by
$\tilde{Q}\in\mathscr{S}^{\prime}$, which will acquire a definite
sense in terms of the modes $\tilde{Q}_{\ell}$. Note the appearance
of the multiplicative term in the momentum operator that depends
both on the measure $\mu_{t_0}$ --uniquely characterized by the
operator $B(t_0)$-- and the operator $A(t_0)$. It guarantees that
the momentum operator is symmetric with respect to the inner
pro\-duct $\langle\cdot|\cdot\rangle_{\mathscr{H}_{s}(t_0)}$.
Indeed, just by using the Gaussian integration by parts formula
$\int_{\mathscr{S}^{\prime}}(D_{\tilde{Q}}\Psi)[f]\,\mathrm{d}\mu_{t_0}[\tilde{Q}]=\int_{\mathscr{S}^{\prime}}\tilde{Q}(\mathcal{C}^{-1}(t_0)f)\Psi[\tilde{Q}]\,\mathrm{d}\mu_{t_0}[\tilde{Q}]$
that can be easily deduced from (\ref{restrict}), we obtain
\begin{eqnarray*}
\big\langle\Phi\,\big|\,\hat{P}_{t_0}[g]\Psi\big\rangle_{\mathscr{H}_{s}(t_0)}&=&i\big\langle
(D_{\tilde{Q}}\Phi)[g]\,\big|\,\Psi\big\rangle_{\mathscr{H}_{s}(t_0)}+i\big\langle\Phi\,\big|\,\tilde{Q}\big(B^{-1}(t_0)(\mathbf{1}+iA(t_0))g\big)\Psi\big\rangle_{\mathscr{H}_{s}(t_0)}\\
&=&i\big\langle(D_{\tilde{Q}}\Phi)[g]+\tilde{Q}\big(B^{-1}(t_0)(\mathbf{1}-iA(t_0))g\big)\Phi\,\big|\,\Psi\big\rangle_{\mathscr{H}_{s}(t_0)}\\
&=&\big\langle\hat{P}_{t_0}[g]\Phi\,\big|\,\Psi\big\rangle_{\mathscr{H}_{s}(t_0)}\,,\,\,\,\forall\,\Phi,\Psi\in\mathscr{D}_{\hat{P}_{t_0}[g]}\,.
\end{eqnarray*}
Let us denote $\hat{Q}_{\ell}(t_{0}):=\hat{Q}_{t_0}[Y_{\ell 0}]$ and
$\hat{P}_{\ell}(t_0):=\hat{P}_{t_0}[Y_{\ell 0}]$, where the
$\hat{Q}_{t_0}[f]$ operator was defined in (\ref{Qoperator}). By
considering the normalization condition (\ref{norm}) and equation
(\ref{invB}), we get
\begin{equation*}
\big(B^{-1}(t_0)(\mathbf{1}-iA(t_0))Y_{\ell
0}\big)(s)=i\frac{\dot{\bar{z}}_{\ell}(t_0)}{\bar{z}_{\ell}(t_0)}Y_{\ell
0}(s)\,,
\end{equation*}
and hence we finally obtain
\begin{equation}\label{QlyPl}
\hat{Q}_{\ell}(t_0)\Psi=\tilde{Q}_{\ell}\Psi\,,\,\,\,\,\hat{P}_{\ell}(t_0)\Psi=-i\frac{\partial\Psi}{\partial
\tilde{Q}_\ell}+\frac{\dot{\bar{z}}_{\ell}(t_0)}{\bar{z}_{\ell}(t_0)}\tilde{Q}_{\ell}\Psi\,,
\end{equation}
where $\Psi$ is a functional of the components $\tilde{Q}_{\ell}$.
The canonical commutation relations
$[\hat{Q}_{\ell}(t_0),\hat{P}_{\ell^{\prime}}(t_0)]=i\delta(\ell,\ell^{\prime})\hat{\mathbb{I}}$
and
$[\hat{Q}_{\ell}(t_0),\hat{Q}_{\ell^{\prime}}(t_0)]=0=[\hat{P}_{\ell}(t_0),\hat{P}_{\ell^{\prime}}(t_0)]$
are obviously satisfied on the appropriate domains.

\bigskip

\indent It is possible to relate the Fock and Schr\"{o}dinger
representations through the action of the annihilation and creation
operators on wave functionals \cite{CorichiSchr1}. Making use of
equations (\ref{alt0}) and (\ref{QlyPl}), we get
\begin{equation}\label{aadag}
\hat{a}_{\ell}(t_0)=\bar{z}_{\ell}(t_0)\frac{\partial}{\partial
\tilde{Q}_{\ell}}\,,\quad\hat{a}_{\ell}^{\dag}(t_0)=-z_{\ell}(t_0)\frac{\partial}{\partial
\tilde{Q}_{\ell}}+\frac{1}{\bar{z}_{\ell}(t_0)}\tilde{Q}_{\ell}\,.
\end{equation}
In particular, the vacuum state is given by the unit constant
functional (up to multiplicative phase)
\begin{equation*}
\Psi^{(t_0)}_{0}[\tilde{Q}]=1\,,\,\,\,\forall\,\tilde{Q}\in\mathscr{S}^{\prime}\,.
\end{equation*}
There exists then a map
$\hat{T}_{t_0}:\mathscr{F}_{\mathcal{P}}\rightarrow\mathscr{H}_{s}(t_0)$
that unitarily connects the creation and annihilation operators of
the Fock and Schr\"{o}dinger representations \cite{Glimm&Jaffe}.
Given the annihilation and creation operators associated with the
modes $z_{\ell}Y_{\ell0}$, $\hat{a}_{\ell}$ and
$\hat{a}_{\ell}^{\dag}$ respectively, the expressions (\ref{aadag})
correspond to
$\hat{T}_{t_0}\circ\hat{a}_{\ell}\circ\hat{T}^{-1}_{t_0}$ and
$\hat{T}_{t_0}\circ\hat{a}_{\ell}^{\dag}\circ\hat{T}^{-1}_{t_0}$,
respectively. These relations, and the action
$\Psi^{(t_0)}_{0}=\hat{T}_{t_0}|0\rangle_{\mathcal{P}}$ on the Fock
vacuum state $|0\rangle_{\mathcal{P}}\in\mathscr{F}_{\mathcal{P}}$,
univocally characterize the unitary transformation $\hat{T}_{t_0}$.
\\
\indent The general procedure that we have followed guarantees the
self-adjointness of the configu\-ration and momentum operators.
Indeed, by the successive action of the creation operator on the
vacuum state $\Psi_{0}^{(t_0)}$, we obtain the $N$-particle states
in the Schr\"{o}dinger representation. These states define, for
$N<\infty$, a common, invariant, dense domain of analytic vectors
for the configuration and momentum operators, so that their
essential self-adjointness is gua\-ranteed, and hence the existence
of unique self-adjoint extensions (see Nelson's analytic vector
theorem in \cite{ReedSimon}).

\bigskip

\indent Finally, the probabilistic interpretation of the models is
given by the usual Born's corres\-pondence rules \cite{Prugovecki}.
Concretely, given $f\in\mathscr{S}$, the theoretical probability
that a measurement carried out in the state $\Psi$ at certain time
to determine the value of $\tilde{Q}(f)$ will yield a result
contained in the Borel set $\Delta\in\mathrm{Bor}(\mathbb{R})$ for
some $\tilde{Q}\in\mathscr{S}^{\prime}$ is given by
\begin{equation}\label{probability}
\mathrm{P}_{\Psi}^{\hat{Q}_{t_0}[f]}(\Delta)=\|\Psi\|_{\mathscr{H}_{s}(t_0)}^{-2}\big\langle\Psi\,\big|\,E^{\hat{Q}_{t_0}[f]}(\Delta)\Psi\big\rangle_{\mathscr{H}_{s}(t_0)}
=\|\Psi\|_{\mathscr{H}_{s}(t_0)}^{-2}\int_{V_{f,\,\Delta}}\big|\Psi[\tilde{Q}]\big|^{2}\,\mathrm{d}\mu_{t_0}[\tilde{Q}]\,,
\end{equation}
where $E^{\hat{Q}_{t_0}[f]}(\Delta)$ is the spectral measure
univocally associated with $\hat{Q}_{t_0}[f]$, defined by
$\big(E^{\hat{Q}_{t_0}[f]}(\Delta)\Psi\big)[\tilde{Q}]=\chi_{V_{f,\,\Delta}}[\tilde{Q}]\,\Psi[\tilde{Q}]$,
with $\chi_{V_{f,\,\Delta}}$ being the characteristic function of
the measu\-rable set
$V_{f,\,\Delta}:=\{\tilde{Q}\in\mathscr{S}^{\prime}\,|\,\tilde{Q}(f)\in\Delta\}\in\sigma(\mathrm{Cyl}(\mathscr{S}^{\prime}))$.
$\|\cdot\|_{\mathscr{H}_{s}(t_0)}$ denotes the norm associated with
the inner product
$\langle\cdot|\cdot\rangle_{\mathscr{H}_{s}(t_0)}$. According to
this, the measure $\mu_{t_0}$ admits the following physical
interpretation: it defines the probability measure
(\ref{probability}) for the vacuum state $\Psi^{(t_0)}_{0}$.

\section{Unitary equivalence of Schr\"{o}dinger representations}\label{UNITARY}

\indent Denote by
$\tau_{(t_0,t_1)}:=\mathfrak{I}_{t_1}^{-1}\circ\mathfrak{I}_{t_0}:\mathbf{P}\rightarrow\mathbf{P}$,
$t_1>t_0$, the symplectomorphism that (i) takes Cauchy data on the
embedding
$\iota_{t_0}(\mathbb{S}^{2})\subset(0,\pi)\times\mathbb{S}^{2}$;
(ii) evolves them to obtain the corres\-ponding solution in
$\mathcal{S}$; and (iii) finally finds the Cauchy data that this
solution induces on the embedding
$\iota_{t_1}(\mathbb{S}^{2})\subset(0,\pi)\times\mathbb{S}^{2}$.
This map implements the classical time evolution from the embedding
$\iota_{t_0}(\mathbb{S}^{2})$ to $\iota_{t_1}(\mathbb{S}^{2})$ on
the canonical phase space, inducing a one-parameter family of states
on the Weyl algebra: Let
$\alpha_{(t_0,t_1)}:\mathcal{A}\rightarrow\mathcal{A}$ be the
$*$-automorphism associated with the symplectic transformation
$\tau_{(t_0,t_1)}$, defined by $\alpha_{(t_0,t_1)}\cdot
W(\lambda):=W(\tau_{(t_0,t_1)}(\lambda))$; the dynamical evolution
of states in the algebraic formulation of the theory is then given
by $\varpi_{t_1}=\varpi_{t_0}\circ\alpha_{(t_0,t_1)}^{-1}$
(Schr\"{o}dinger picture), with $\varpi_{t_0}$ defined in equation
(\ref{statesigma}). The evolved state $\varpi_{t_1}$ acts on the
elementary observables as
$\varpi_{t_1}(W(\lambda))=\exp\big(-\omega(J_{t_1}(\lambda),\lambda)/4\big)$,
where the complex structure
\begin{equation*}
J_{t_{1}}:=\tau_{(t_0,t_1)}\circ
J_{t_0}\circ\tau^{-1}_{(t_0,t_1)}=\mathfrak{I}_{t_1}^{-1}\circ
J_{\mathcal{P}}\circ\mathfrak{I}_{t_1}:\mathbf{P}\rightarrow\mathbf{P}
\end{equation*}
defines a new Schr\"{o}dinger representation\footnote{Here, we will
make a notational abuse and simply denote the triplet
$\big(\mathscr{H}_{s}(t),\pi_{s}(t),\Psi^{(t)}_{0}\big)$ as
$\mathscr{H}_{s}(t)$.} $\mathscr{H}_{s}(t_1)$. Clearly, the
condition of unitary equivalence of the Schr\"{o}dinger
representations corresponding to different values $t_0<t_1$ of the
time parameter amounts to demanding the unitary implementability of
the symplectic transformation $\tau_{(t_0,t_1)}$ in the
$\mathscr{H}_{s}(t_0)$ representation\footnote{In this way,
$J_{t_1}-J_{t_0}$ is a Hilbert-Schmidt operator in the one-particle
Hilbert space constructed from $J_{t_0}$ (or equivalently
$J_{t_1}$).}. In that case, there exists a unitary transformation
$\hat{V}_{(t_0,t_1)}:\mathscr{H}_{s}(t_0)\rightarrow\mathscr{H}_{s}(t_1)$
mapping the configuration and momentum operators from one
representation into the other, in such a way that
\begin{eqnarray}\label{Vt0t1}
\hat{V}_{(t_0,t_1)}\circ\hat{a}_{\ell}(t_0)\circ\hat{V}^{-1}_{(t_0,t_1)}&=&\alpha_{\ell}(t_0,t_1)\hat{a}_{\ell}(t_1)+\beta_{\ell}(t_0,t_1)\hat{a}^{\dag}_{\ell}(t_1)\,,\\
\hat{V}_{(t_0,t_1)}\circ\hat{a}^{\dag}_{\ell}(t_0)\circ\hat{V}^{-1}_{(t_0,t_1)}&=&\bar{\beta}_{\ell}(t_0,t_1)\hat{a}_{\ell}(t_1)+\bar{\alpha}_{\ell}(t_0,t_1)\hat{a}^{\dag}_{\ell}(t_1)\,,\nonumber
\end{eqnarray}
where
\begin{equation}\label{alpha&beta}
\alpha_{\ell}(t_0,t_1):=i\Big(\bar{z}_{\ell}(t_0)\dot{z}_{\ell}(t_1)-z_{\ell}(t_1)\dot{\bar{z}}_{\ell}(t_0)\Big),\,\,\,\,
\beta_{\ell}(t_0,t_1):=i\Big(\bar{z}_{\ell}(t_0)\dot{\bar{z}}_{\ell}(t_1)-\bar{z}_{\ell}(t_1)\dot{\bar{z}}_{\ell}(t_0)\Big).
\end{equation}
According to the results achieved in \cite{BarberoG.:2007qt}, once
we consider an $SO(3)$ invariant complex structure $J_{\mathcal{P}}$
verifying the conditions (\ref{rhonu}), the quantum dynamics can be
unita\-rily implemented in $\mathscr{F}_{\mathcal{P}}$, i.e., there
exists a unitary operator
$\hat{U}(t,t^{\prime}):\mathscr{F}_{\mathcal{P}}\rightarrow\mathscr{F}_{\mathcal{P}}$
encoding the information about the evolution of the system from time
$t$ to $t^{\prime}$. This condition is precisely ensured by the
square summability of the $\beta_{\ell}$ coefficients appearing in
the Bogoliubov transformation (\ref{Vt0t1}), and guarantees that the
map $\hat{V}_{(t_0,t_1)}$ is well defined, i.e., the Schr\"{o}dinger
representations corresponding to different times $t_0,t_1$ are
equivalent. The unitary transformation
$\hat{V}_{(t_0,t_1)}=\hat{T}_{t_1}\circ\hat{U}(t_0,t_1)\circ\hat{T}_{t_0}^{-1}$
relating them is completely cha\-racterized by the relations
(\ref{Vt0t1}) and the action on the vacuum state
$\Psi^{(t_0)}_{0}\in\mathscr{H}_{s}(t_0)$, given by
\begin{equation}\label{VPsi0}
\Big(\hat{V}_{(t_0,t_1)}\Psi^{(t_0)}_{0}\Big)[\tilde{Q}]=\prod_{\ell=0}^{\infty}\frac{|z_{\ell}(t_1)|^{1/2}}{|z_{\ell}(t_0)|^{1/2}}\exp\left(-\frac{1}{2}\frac{\beta_{\ell}(t_0,t_1)}{\bar{z}_{\ell}(t_0)\bar{z}_{\ell}(t_1)}\tilde{Q}_{\ell}^{2}\right)\in\mathscr{H}_{s}(t_1)\,,
\end{equation}
where we have used the fact that
$\hat{a}_{\ell}(t_0)\Psi^{(t_0)}_{0}=0$,
$\forall\,\ell\in\mathbb{N}\cup\{0\}$, and the expressions
(\ref{norm}), (\ref{aadag}) and (\ref{Vt0t1}) to obtain the
differential equations verified by this state; namely, $\partial
\hat{V}_{(t_0,t_1)}\Psi^{(t_0)}_{0}/\partial\tilde{Q}_{\ell}=-\big(\beta_{\ell}(t_0,t_1)/\bar{z}_{\ell}(t_0)\bar{z}_{\ell}(t_1)\big)\tilde{Q}_{\ell}\hat{V}_{(t_0,t_1)}\Psi^{(t_0)}_{0}$,
$\ell\in\mathbb{N}\cup\{0\}$. The equation (\ref{VPsi0}) must be
interpreted as the limit in the $\mathscr{H}_{s}(t_1)$-norm of the
Cauchy sequence of normalized vectors $f_{n}\in\mathscr{H}_{s}(t_1)$
obtained by extending the product (\ref{VPsi0}) to a finite integer
$n\in\mathbb{N}$.

\bigskip

\indent The mutual absolute continuity of any two Gaussian measures
associated with different times $t_0,t_1\in(0,\pi)$ is also
verified\footnote{It is possible to show that the equivalence of
measures is a necessary condition for the unitary equivalence
between Schr\"{o}dinger representations \cite{Mena:2007schr}.}, i.e.
they have the same zero measure sets. This property requires that
the operator $\mathcal{C}(t_1)-\mathcal{C}(t_0)$ is Hilbert-Schmidt
\cite{Simon,Yamasaki,Mourao}, which is satisfied in our case.
Indeed, it is straightforward to check that the sequence
$\big(|z_{\ell}(t_1)|^{2}-|z_{\ell}(t_0)|^2\big)_{\ell=0}^{\infty}$
is square summable. On the contrary, for the original scalar field
$\phi=\xi/\sqrt{\sin t}$, for which the time evolution is not
unitary, we get the nonequivalence of the representations obtained
for different times, and also the impossibility of such continuity.
In this case, the mutual singularity of measures can be expected, as
was proved for the vacuum Gowdy $\mathbb{T}^{3}$ model in
\cite{Torre:2007zj}.

\bigskip

\indent Note that the map
$\hat{T}_{t_0}:\mathscr{F}_{\mathcal{P}}\rightarrow\mathscr{H}_{s}(t_0)$
introduced in subsection \ref{CCR} does not connect the
configuration and momentum operators of the Fock representation,
$\hat{Q}_{\ell}(t)=z_{\ell}(t)\hat{a}_{\ell}+\bar{z}_{\ell}(t)\hat{a}_{\ell}^{\dag}$
and
$\hat{P}_{\ell}(t)=\dot{z}_{\ell}(t)\hat{a}_{\ell}+\dot{\bar{z}}_{\ell}(t)\hat{a}_{\ell}^{\dag}$,
respectively, with those of the Schr\"{o}dinger one (except for
$t=t_0$). However, owing to the unitary implementability of the
dynamics, there exists also a unitary transformation
$\hat{V}_{\mathscr{F}_{\mathcal{P}},t_0}(t):\mathscr{F}_{\mathcal{P}}\rightarrow\mathscr{H}_{s}(t_0)$,
such that $\hat{V}_{\mathscr{F}_{\mathcal{P}},t_0}(t)\circ
\hat{a}_{\ell}\circ\hat{V}_{\mathscr{F}_{\mathcal{P}},t_0}^{-1}(t)=\alpha_{\ell}(t,t_0)\hat{a}_{\ell}(t_0)+\beta_{\ell}(t,t_0)\hat{a}_{\ell}^{\dag}(t_0)$,
$\hat{V}_{\mathscr{F}_{\mathcal{P}},t_0}(t)\circ
\hat{a}^{\dag}_{\ell}\circ\hat{V}_{\mathscr{F}_{\mathcal{P}},t_0}^{-1}(t)=\bar{\beta}_{\ell}(t,t_0)\hat{a}_{\ell}(t_0)+\bar{\alpha}_{\ell}(t,t_0)\hat{a}_{\ell}^{\dag}(t_0)$,
relating these operators. In terms of the unitary evolution operator
on $\mathscr{F}_{\mathcal{P}}$, we have
$\hat{V}_{\mathscr{F}_{\mathcal{P}},t_0}(t)=\hat{T}_{t_0}\circ\hat{U}^{-1}(t_0,t)$.
Fina\-lly, given the quantum Hamiltonian (\ref{QuantHamiltonian}) in
the Fock representation, with dense domain
$\mathscr{D}_{\hat{H}_{0}(t)}\subset\mathscr{F}_{\mathcal{P}}$
spanned by the states with a finite number of particles, the
corresponding operator in the $\mathscr{H}_{s}(t_0)$ representation
is given by
$\hat{V}_{\mathscr{F}_{\mathcal{P}},t_0}(t)\circ\hat{H}_{0}(t)\circ\hat{V}_{\mathscr{F}_{\mathcal{P}},t_0}^{-1}(t)$,
\begin{equation*}\label{SchquantHamilt}
\frac{1}{2}\sum_{\ell=0}^{\infty}\Bigg[-\frac{\partial^{2}}{\partial\tilde{Q}_{\ell}^{2}}
-2i\frac{\dot{\bar{z}}_{\ell}(t_0)}{\bar{z}_{\ell}(t_0)}\tilde{Q}_{\ell}\frac{\partial}{\partial\tilde{Q}_{\ell}}
+\left(\frac{\dot{\bar{z}}^{2}_{\ell}(t_0)}{\bar{z}_{\ell}^{2}(t_0)}+\frac{1}{4}\big(1+\csc^{2}t\big)+\ell(\ell+1)\right)\big(\tilde{Q}_{\ell}^{2}-|z_{\ell}(t_0)|^{2}\big)\Bigg]
\end{equation*}
modulo an irrelevant real term proportional to the identity. Note,
by contrast, that the complex independent term appearing in the
previous expression is necessary to ensure that the operator is
self-adjoint. This Hamiltonian is defined in the dense subspace
$\hat{V}_{\mathscr{F}_{\mathcal{P}},t_0}(t)\mathscr{D}_{\hat{H}_{0}(t)}=\big\{\hat{V}_{\mathscr{F}_{\mathcal{P}},t_0}(t)f\,|\,f\in\mathscr{D}_{\hat{H}_{0}(t)}\big\}\subset\mathscr{H}_{s}(t_0)$
generated by the cyclic vector
$\hat{V}_{\mathscr{F}_{\mathcal{P}},t_0}(t)|0\rangle_{\mathcal{P}}\in\mathscr{H}_{s}(t_0)$.

\section{Comments}\label{COMMENTS}

\indent We have constructed the Schr\"{o}dinger representation for
the linearly polarized Gowdy $\mathbb{S}^{1}\times\mathbb{S}^{2}$
and $\mathbb{S}^{3}$ models coupled to massless scalar fields in a
mathematically rigorous and self-contained way, completing in this
way the quantization of these systems given in
\cite{BarberoG.:2007qt}. We have assumed the use of the redefined
fields for which the dynamics is well defined and unitary. As proved
in \cite{BarberoG.:2007qt,MenaSchr}, the complex structures
$J_{\mathcal{P}}$ verifying the conditions (\ref{rhonu}) lead to
unitarily equivalent quantum theories, and hence the Schr\"{o}dinger
representations corresponding to them are also equivalent. Note
that, as far as the support of the measure or the unitary
implementability of the dynamics is concerned, the discussions and
results obtained for these models are analogous to those found for
the vacuum $\mathbb{T}^{3}$ model in \cite{Torre:2007zj} and
\cite{Mena:2007schr}. It could be argued that this similarity is
somehow expected due to the fact that the critical features of the
systems are determined by their ultraviolet behaviors, and these
should not be sensitive to the topology of the spacetimes. This
argument can be found, for example, in \cite{Birrell&Davies}
concerning the simplest generalization of Minkowski space quantum
field theory to the $\mathbb{R}\times\mathbb{T}^{3}$ spacetime with
closed spatial sections. This compactification can modify the
long-wavelength behavior of the system, but not the ultraviolet one,
so that both spacetimes suffer from the same ultraviolet divergence
properties. Such statement is clearly intuitive, but it is not
obvious to what extent it is true for quantum field theories in
spacetimes, like those corresponding to the Gowdy models, that are
not locally isometric. In this respect, the similarity of the
results is probably due to the similar structure of the differential
equations verified by the mode functions. In any case, it is
interesting to analyze in a rigorous way the particularities of the
quantizations for the different topologies.

\bigskip

\indent Finally, it is important to highlight the advantage of using
the re-scaled fields that make the quantum dynamics unitary, given
that in this case it is possible to obtain a unique (up to unitary
equivalence) Fock/Schr\"{o}dinger representation for these models.
As a direct consequence, the mutual absolute continuity of the
measures corresponding to different times is verified. Neither of
these properties can be attained for the original variables. In this
situation, even if the failure of the unitarity of time evolution
and the mutual singularity of measures are not serious obstacles for
a suitable probabilistic interpretation of the models
\cite{Torre:2002xt,Torre:2007zj}, we must face the lack of
uniqueness of the representation.

\begin{acknowledgments}
The author is indebted to J. Fernando Barbero G. and Eduardo J. S.
Villase\~{n}or for many enlightening discussions and helpful
suggestions. He also wishes to thank G. A. Mena Marug\'an for his
valuable comments regarding the uniqueness of the Fock
representation, that have led to include some necessary
clarifications in the main body of the paper. The author
acknowledges the support of the Spanish Research Council (CSIC)
through a I3P research assistantship. This work is also supported by
the Spanish MEC under the research grant FIS2005-05736-C03-02.

\end{acknowledgments}

\end{document}